# Flexible A-scan rate MHz OCT:
# Computational downscaling by coherent averaging

Tom Pfeiffer[1], Wolfgang Wieser[3], Thomas Klein[3], Markus Petermann[2,3], Jan-Phillip Kolb[1], Matthias Eibl[1] and Robert Huber[1,3]

[1]Institut für Biomedizinische Optik, Universität zu Lübeck, Peter-Monnik-Weg 4, 23562 Lubeck, Germany
[2]Lehrstuhl für BioMolekulare Optik, Fakultät für Physik, Ludwig-Maximilians-Universität München, Oettingenstr. 67, 80538 Munich, Germany
[3]Optores GmbH, Gollierstr. 70, 80339 Munich, Germany

## ABSTRACT

In order to realize fast OCT-systems with adjustable line rate, we investigate averaging of image data from an FDML based MHz-OCT-system. The line rate can be reduced in software and traded in for increased system sensitivity and image quality. We compare coherent and incoherent averaging to effectively scale down the system speed of a 3.2 MHz FDML OCT system to around 100 kHz in postprocessing. We demonstrate that coherent averaging is possible with MHz systems without special interferometer designs or digital phase stabilisation. We show OCT images of a human finger knuckle joint in vivo with very high quality and deep penetration.

## 1. INTRODUCTION

Modern FDML lasers have paved the way to high fidelity swept source OCT systems with A-scan rates up to several MHz [1] and coherence lengths superior to usual spectrometer based systems [2, 3]. Today also other technologies can achieve MHz line rates [4-6]. While MHz-OCT enables many new applications such as wide field ophthalmic OCT [7-10], flicker free live 4D-OCT [11] and high resolution intravascular OCT without motion artifacts [12, 13], the short acquisition times always come at cost of system sensitivity. But there are clearly situations when image quality is more important than acquisition speed. Ideally the same OCT system could perform high speed imaging with reduced sensitivity and high sensitivity imaging at lower speed. Sadly the hardware of most high speed swept source OCT systems cannot run at arbitrary line rates, even though there are light sources with adjustable wavelength sweep repetition rate [14-17]. The reason is that the high performance optoelectronic periphery of modern OCT engines often needs to be specifically tailored to a certain line rate. In most real OCT systems analog electronic filters, the bandwidth of photo-receivers, the mechanical response of resonant galvanometer-scanners and much more has to be designed for a certain line rate to achieve shot noise limited sensitivity. An alternative approach to realize variable line rate OCT is to image fast and reduce the line rate afterwards in software by averaging, which also improves image quality. The ability of coherent averaging to reduce white Gaussian noise more effectively than incoherent averaging has already been demonstrated on OCT data [18, 19]. Only coherent averaging, which comprises averaging of fringes or the complex output of the FFT, can improve sensitivity in the same way as lower imaging speed does. Now on the one hand, coherent averaging should be highly compatible with FDML based MHz-OCT since FDML lasers exhibit very good phase stability and the high speed effectively suppresses negative effects by sample motion. On the other hand, the low transimpedance gain of MHz-OCT photodetectors increases the negative effect of non-Gaussian correlated noises such as excess laser and ADC noise on the averaging result. These non-Gaussian noise contributions are more resistant to complex averaging. Hence, white noises such as shot noise or Johnson noise are not necessarily the limiting factors for effective averaging in FDML based MHz OCT systems. So this paper investigates downscaling of A-scan rates in MHz-OCT by coherent averaging.

## 2. METHODS

All data shown has been acquired using a home-built 3.2 MHz FDML laser and the stability of the interferometer was not optimized. The real system sensitivity was measured to be 106 dB with an optical power of 40 mW on the sample. In one imaging experiment the sensitivity was artificially reduced to 70 dB using an additional ND filter. In the following we will denote the averaging of complex FFT outputs as coherent averaging and averaging of the absolute values of the FFT outputs as incoherent averaging.

## 3. RESULTS AND DISCUSSION

To estimate an upper limit for the efficiency of coherent averaging we measured the decrease of the background noise over the number of a-scans averaged. Figure 1 illustrates the decrease of the background standard deviation due to coherent averaging. As can be read from the graph, coherent averaging efficiently reduces the background noise.

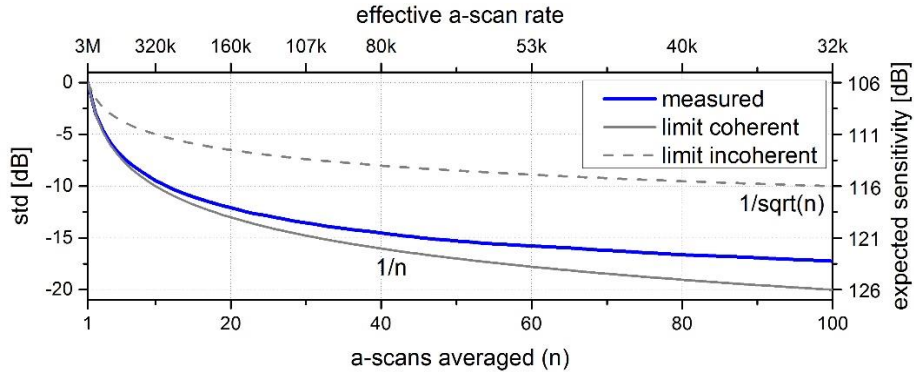

**Figure 1:** Decrease of the background noise due to coherent averaging depending on the number of A-scans averaged. The data was acquired with blocked sample arm. Within one frame adjacent A-scans were averaged coherently and the background noise was measured within a depth area corresponding to the medium third of the averaged A-scan. The blue curve is the average of hundred curves acquired in this way while the gray curve indicates the theoretical limit for coherent averaging and the gray dashed line indicates the theoretical limit for incoherent averaging.

However there is an deviation from the theoretical limit that increases with the number of A-scans averaged and is plotted in Figure 2. This averaging penalty is probably due to excess noise or ADC related noise becoming dominant. The spectral components of these noise sources do not have random phase distributions and are therefore more resistive to coherent averaging. As can be read from the graph, we should be able to downscale the a-scan rate of our system by a factor of 30 with a resulting sensitivity 1.25dB less than that of an comparable ideal 107 kHz system. This of course is only true assuming that the signal isn't degraded due to A-scan to A-scan phase variations.

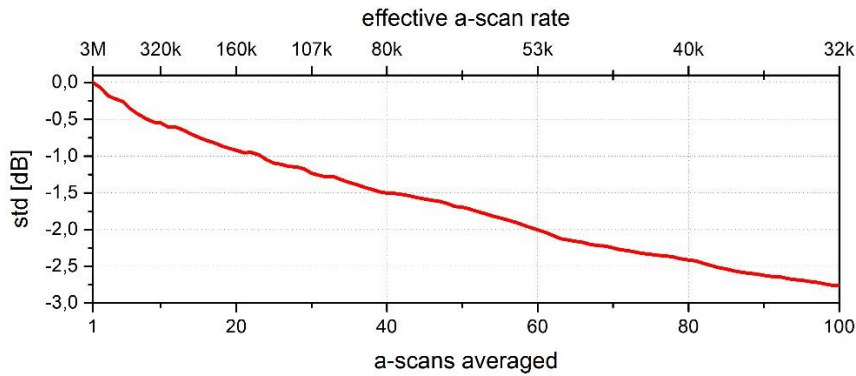

Figure 2: The red line indicates the deviance of the measured decrease in background noise in **Figure 1** to the theoretical limit for coherent averaging.

To verify the benefit of coherent averaging we imaged a human finger knuckle joint covered with ultrasound gel pressed against a cover glass. **Figure 3** gives a comparison of the effects of coherent and incoherent averaging, downscaling the system speed to an effective a-scan rate of 107 kHz. To eliminate the speckle pattern 15 slightly displaced images were averaged incoherently. It should be noted that the cutlevels where set differently for the two resulting images. This is necessary for a fair comparison of the two averaging modalities because coherent averaging actually reduces the background instead of just better defining it as incoherent averaging does. We preferred this strategy to histogram equalization since the visual impression in the regions with signal is more comparable this way. Both averaging techniques vastly improve the image quality, though the difference between the two is not as dramatic as may be expected. While we clearly see improved contrast due to the reduced uneven background in the case of coherent averaging, improved sensitivity, i.e. in this case seeing features which cannot be identified with the other technique, cannot be claimed.

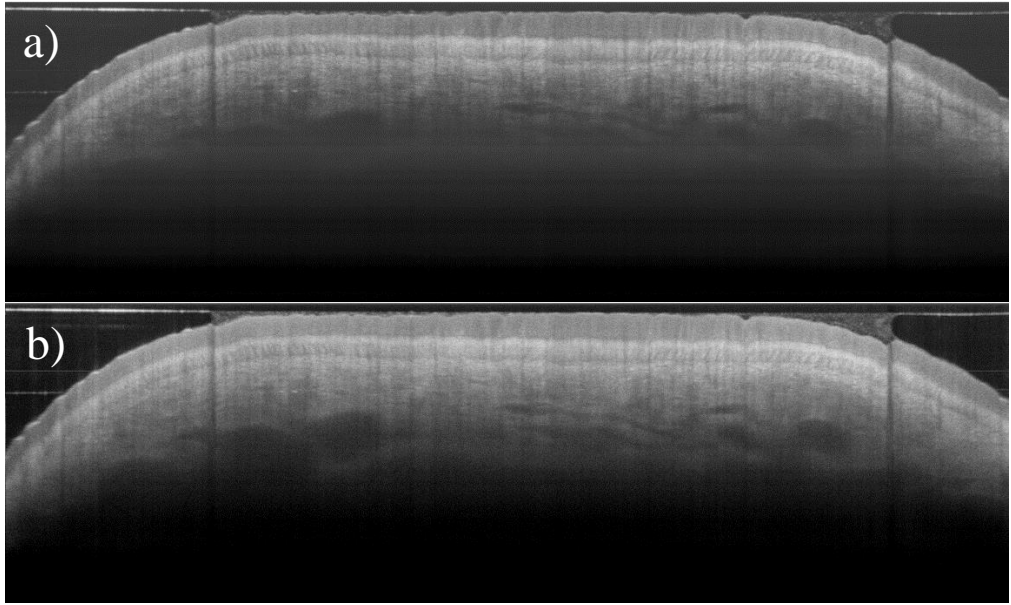

**Figure 3:** OCT images of human finger knuckle joint. The same dataset acquired with 3.2MHz OCT. The dataset consists of 15 frames with 60000 a-scans each. **a)** The a-scans were averaged **incoherently** in groups of 30 to form a single a-scan. All 15 frames were averaged incoherently. **b)** A-scans were **coherently** averaged to form a single a-scan in groups of 30. All 15 frames were then averaged incoherently.

To examine the effect of coherent averaging on low sensitivity systems we reduced the system sensitivity to 70dB by attenuating the sample arm with an OD 1.8 neutral density filter and acquired images of a piece of orange as a stationary sample. **Figure 4** compares the unaveraged result to a 30 times incoherently and a 30 times coherently averaged image. In this case the difference between the two averaging modalities is much more dramatic. While both techniques improve the image quality, the coherently averaged image reveals much more details and the uneven background that is most distinct in the middle third of the image is vastly reduced. Again, the cutlevels have been adjusted for the higher dynamic range in the coherently averaged case. The clearly visible advantages of coherent averaging indicate that it might be esapecially usefull to downscale the a-scan rate of sensitivity limited systems. The most obvious candidate might be retinal imaging systems, where the applicable sample arm power is limited.

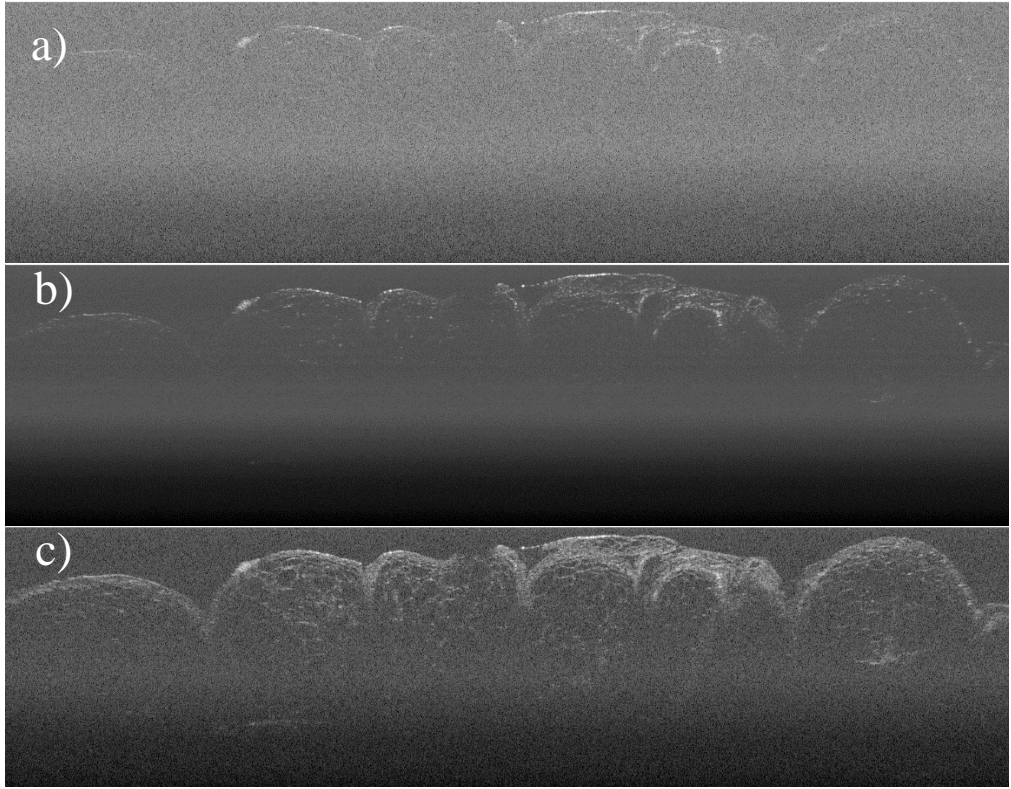

**Figure 4:** Orange , imaged with -36dB attenuation. The dataset is one frame with 60000 a-scan. **a)** An image without any averaging was constructed stitching together every 30th a-scan to compase a frame **b)** The a-scans were averaged incoherently in groups of 30 to form a single A-scan. **c)** A-scans were coherently averaged to form a-single a-scan in groups of 30.

## 4. CONCLUSION AND OUTLOOK

We have investigated the feasibility of coherent averaging of A-scans using a MHz FDML OCT system. We observe that coherent computational downscaling of a MHz OCT system is possible and useful up to a certain point. Especially "low sensitivity" OCT systems might significantly benefit from complex averaging, whereas systems already starting at ~106dB exhibit less improvement. It is therefore possible to demonstrate very high quality images acquired with a 3.2 MHz OCT. Our results indicate that the computational downscaling in software might especially be compatible with retinal MHz-OCT systems since they only have ~90dB system sensitivity.